\documentclass[aps,pra,reprint,floatfix,superscriptaddress,amssymb,amsmath]{revtex4-1}
\usepackage{graphicx}
\usepackage{hyperref}
\usepackage{enumitem}
\usepackage{braket}
\usepackage{color}
\usepackage{amsthm}
\usepackage{bm} % For bold greek letters

\newcommand{\diff}{\mathop{}\!\mathrm{d}}
\newcommand{\comma}{\text{,}}
\newcommand{\semicomma}{\text{;}}
\newcommand{\point}{\text{.}}

\newcommand{\vc}[1]{{\bm{#1}}}
\newcommand{\angstrom}{\textup{\AA}}

\begin{document}

\title{Quantum Emulation of Molecular Force Fields: A Blueprint for a Superconducting Architecture}

\author{Diego Gonz\'{a}lez \surname{Olivares}}
\affiliation{Instituto de F\'{i}sica Fundamental IFF-CSIC, Calle Serrano 113b, Madrid E-28006, Spain.}
\author{Borja \surname{Peropadre}}
\affiliation{Quantum Information Processing group, Raytheon BBN Technologies, 10 Moulton Street, Cambridge, Massachusetts 02138, United States}
\author{Joonsuk \surname{Huh}}
\affiliation{Department of Chemistry, Sungkyunkwan University, Suwon 440-746, Korea}
\author{Juan Jos\'e \surname{Garc\'{i}a-Ripoll}}
\affiliation{Instituto de F\'{i}sica Fundamental IFF-CSIC, Calle Serrano 113b, E-28006 Madrid, Spain.}
 
\begin{abstract}
In this work, we propose a flexible architecture of microwave resonators with tunable couplings to perform quantum simulations of problems from the field of molecular chemistry. The architecture builds on the experience of the D-Wave design, working with nearly harmonic circuits instead of qubits. This architecture, or modifications of it, can be used to emulate molecular processes such as vibronic transitions. Furthermore, we discuss several aspects of these emulations, such as dynamical ranges of the physical parameters, quenching times necessary for diabaticity, and, finally, the possibility of implementing anharmonic corrections to the force fields by exploiting certain nonlinear features of superconducting devices.
\end{abstract}

\maketitle

\section{Introduction}
\label{sec:Introduction}

Among the different controllable quantum systems in the field of quantum technologies, superconducting circuits excel at the possibility of establishing interconnected scalable architectures and tunable long-range couplings. A paradigmatic example of this possibility is the D-Wave architecture~\cite{bunyk14}, in which arrays of superconducting flux qubits are controlled with various parameters: qubit frequencies, biases, and coupling strengths between nearest neighbors and connectivities to other plaquettes~\cite{harris09,johnson10}. All of these advances were achieved in a recent implementation of adiabatic quantum optimizers (also known as quantum annealers)~\cite{boixo14}. Concerning the architecture, the focus is currently placed on improving the quality of qubits, increasing their coherence times and pushing towards larger system sizes.

In this work, we explore alternative routes where the capabilities of D-Wave--like superconducting circuits are leveraged in the context of less demanding applications in quantum simulation. The key idea is that the D-Wave circuit can be moved into a regime in which it behaves as a collection of thousands of resonators with tunable frequencies, couplings, and nonlinearities. It is a powerful platform that can be used to study a wide variety of models. We show how it can be used to emulate the molecular force fields that govern the vibrational dynamics of complex molecules, with or without anharmonicities.

\begin{figure}[ht!]
\centering
\includegraphics[width = 0.725\linewidth]{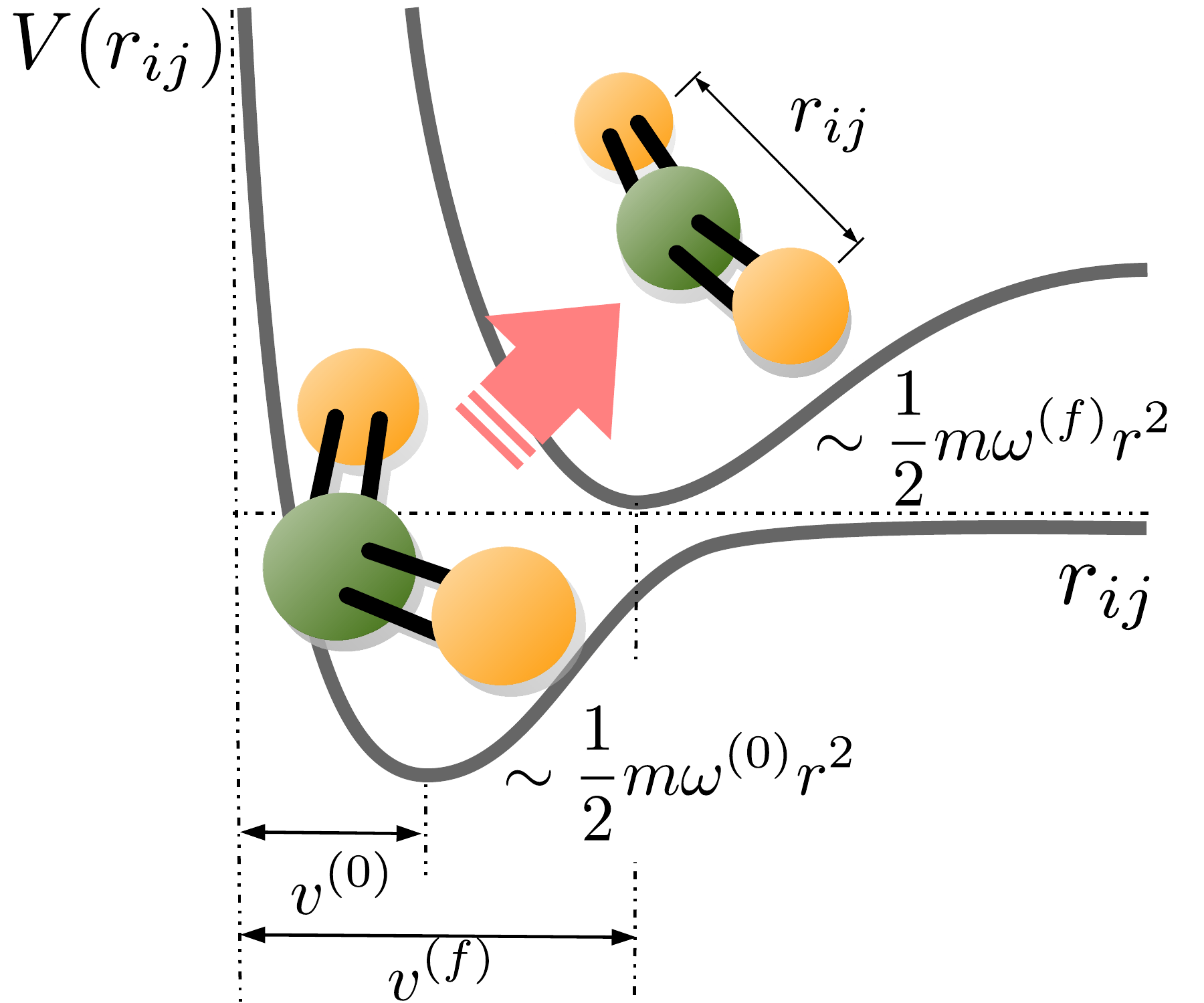}
\caption{When a molecule is electronically excited, it experiences a sudden change in its force field. This change leads to an effective quench which excites the phonon degrees of freedom. After this quench, the molecule may relax to the new ground state by releasing its excess energy.}
\label{fig:the-problem}
\end{figure}

The problem we have in mind is sketched in Fig.~\ref{fig:the-problem}, which depicts a molecule that undergoes a change in its force field as a consequence of an electronic transition. Because of the molecular restructuring, the vibrational modes are displaced, mixed, and squeezed through a Duschinsky rotation, which makes computing the energy distribution for these modes after a sudden transition ---the Franck-Condon Profile---, a challenging problem~\cite{huh14}, even more so if anharmonicities are considered~\cite{huh:2010anharm}.

Here, we show how to map the vibrational structure of a molecule to a superconducting circuit emulator. This map allows us to imitate interatomic interactions, emulate sudden, adiabatic, or intermediate quenches, and experimentally reproduce the Franck-Condon profile in the harmonic regime---i.e., within a quadratic approximation to the force field potential---with the possibility hinted at below of including anharmonic corrections to the force field.

Our proposal is complementary to other applications and proposals for using superconducting circuits~\cite{omalley15} and other quantum architectures~\cite{Shen17} to obtain answers to different questions posed in the realm of molecular physics and quantum chemistry, such as the study of ground-state properties of certain molecules~\cite{aspuru-guzik05} or transport phenomena~\cite{garcia-alvarez15}. It has already been shown that cavity arrays with qubits and boson-sampling techniques~\cite{huh14} can provide information about molecular vibrational spectra and that they may be implemented using superconducting circuits~\cite{peropadre15}. We now propose using the same building blocks in a direct implementation of the molecular Hamiltonian, which provides the possibility of taking into account finite quenching speeds (beyond the adiabatic and Franck-Condon approximations), and the effects of other, possibly time-dependent, fields.

This work is structured in three parts. We begin in Sec.~\ref{sec:the-problem} by introducing the particular problem that we intend to emulate: the computation of the Franck-Condon profile~\cite{huh14} that describes the distribution of energies in the vibrational degrees of freedom of a molecule after a sudden change of its force field. In Sec.~\ref{sec:protocol} we discuss, from a theoretical point of view, which operations and protocols are needed to simulate the force field and to reproduce the Franck-Condon profile. We produce mathematically rigorous bounds for a possible physical implementation of the emulator, including preparation, quenching times, and measurement protocols. Finally, Sec.~\ref{sec:Architecture} provides one possible physical implementation of the emulator using a multiply connected architecture of tunable microwave resonators, together with some qubits for measurement purposes. This section leverages on our understanding of such architectures from the realm of circuit QED and quantum annealers, and it shows the feasibility of implementing hundreds of vibrational modes together with detailed controls and measurement schemes. We conclude this work with a discussion of possible avenues where these simulators may provide alternative physical insights, such as the study of nonlinear terms in the molecular force field (see, e.g., Ref.~\cite{huh:2010anharm}), or applications to other computational problems, such as spin models with long-range interactions~\cite{peropadre15,Ortiz14}.

\section{A molecular problem: Franck-Condon profile}
\label{sec:the-problem}

When a photoinduced electronic transition takes place in a molecule, it experiences vibrational transitions along with the electronic transition (see, e.g., Refs.~\cite{Wilson,Atkins,Huh2011a}). The molecular process is usually described in terms of the Born-Oppenheimer potential energy surface $V\left(\vc{x}_1\comma\ldots\comma\,\vc{x}_N\semicomma\,\Theta\right)$ of the electronic configuration $\ket{\Theta}$, which is a function of the atomic positions $\left\lbrace\vc{x}_{i}\right\rbrace$. Here, the state of the molecule is described by a vibronic wave function $\ket{\Psi}\otimes\ket{\Theta}$, with separable vibrational ($\ket{\Psi}$) and electronic ($\ket{\Theta}$) components. For a given electronic configuration $\ket{\Theta}$, the dynamics of the nuclei in the molecule can be approximated as a set of coupled harmonic oscillators,
\begin{equation}
\label{eq:Hamiltonian}
H_\Theta = \sum_j \frac{1}{2m_j} \, \vc{p}_j^2
         + \frac{1}{2} \sum\limits_{jk} \left( \vc{x}_j - \vc{v}_j \right)^T \hspace{-0.3em}
                                        \hat{A}^{\left(\Theta\right)}_{jk}
                                        \left( \vc{x}_k - \vc{v}_k \right) \point
\end{equation}
Here, $\vc{v}_j$ represents the equilibrium positions of the $j$th atom and $\hat{A}^{\left(\Theta\right)}$ is the quadratic force constant matrix (Hessian). $m_{j}$ and $\vc{p}_{j}$ are, respectively, the mass and the momentum corresponding to the $j$th atom. A Hamiltonian of this form describing the mechanical vibrations of the atoms in a molecule also has an implicit physical symmetry, which are the translational and rotational invariances associated with the rigid motions of the molecule.

With the additional assumption (the Condon approximation) of considering the electronic transition moment $d_{\left(0,f\right)} = \bra{\Theta_f} \sum_j e \, \vc{r}_j \ket{\Theta_0}$ (where the vectors $\vc{r}_j$ are the electronic positions) to be independent of the nuclear positions $\vc{x}_j$, what we are describing here is a sudden electronic transition from an initial vibronic state $\ket{\Psi_0}\otimes\ket{\Theta_0}$ to a final state $\ket{\Psi_f}\otimes\ket{\Theta_f}$. As a consequence of the dipole moment $d_{\left(0,f\right)}$ being constant, the transition amplitudes for the nuclear wave function $\braket{\Psi_f\vert\Psi_0}$ characterize the transition profile.

After the electronic transition, the nuclei experience a quantum quench, abruptly experiencing a different force field with a different Hamiltonian [still of the form of Eq.~\eqref{eq:Hamiltonian}]. The molecule undergoes a structural deformation as the initial wave function $\ket{\Psi_0}$ evolves on a new potential energy surface. As the final nuclear Hamiltonian $H_{\Theta_f}$ can also be approximated by a quadratic interaction Hamiltonian of the form of Eq.~\eqref{eq:Hamiltonian}, it will also have occupation-number eigenstates $\ket{\vc{n}}$ and eigenenergies $E_{\vc{n}} = \sum_j \, \omega_j n_j$. Consequently, the state of the system will evolve as
\begin{align} \label{eq:FC-State}
\ket{\Psi\left(t\right)} & = e^{-i \, \left(H_{\Theta_f}+E_{\mathrm{el}}\right) \, t} \ket{\Psi_0} \nonumber \\
                         & = \sum\limits_\vc{n} \braket{ \vc{n} \vert \Psi_0 }
                                                e^{-i(E_\vc{n}+E_{\mathrm{el}})t} \ket{\vc{n}} \comma
\end{align}
where $E_{\mathrm{el}}$ is the adiabatic electronic transition energy; this offset energy can be safely set to zero in our description of the transition probability distribution. The probability distribution $P\left(E\right)$ associated with finding the above nonequilibrium states in a given energy manifold is called the Franck-Condon profile~\cite{jankowiak:2007}:
\begin{align} \label{eq:Franck-Condon-profile}
P\left(E\right) = \sum\limits_\vc{n} \delta \left( E - E_\vc{n} \right)
                  \left\vert \braket{ \vc{n} \vert \Psi_0 } \right\vert^2 \point
\end{align}

The problem with classically reproducing the Franck-Condon profile, the main physical observable pertaining these transitions that can be obtained experimentally, is linked to two different layers of difficulty. The first one consists of the fact that reconstructing $P\left(E\right)$ involves sampling subsets of instances with specific population numbers and is thereby related to the problem of boson sampling, which is conjectured to be challenging~\cite{huh14,rahimi2015,huh2016}. More precisely, the original boson sampling problem~\cite{Aaronson2011} can appear in the current molecular picture by making all normal mode frequencies approximately the same, preparing the initial Fock states and studying the evolution under an arbitrarily complex change of the force field. The second difficulty lies in the relationship between $P\left(E\right)$ and the integer partition problem. Simply put, a given energy $E = \sum_j \omega_j \, n_j$ may be approximated by many different sets of phonon occupation numbers, and reconstructing the full probability distribution involves counting every such possible configuration and calculating all of the possible overlaps in Eq.~\eqref{eq:Franck-Condon-profile}~\cite{huh14}. Arguably, the two problems combined make the classical estimation of $P\left(E\right)$ difficult under general circumstances.

\section{Quantum emulation protocol}
\label{sec:protocol}

In this section, we analyze the steps necessary for implementing a complete emulation of the molecular dynamics during a quantum quench of its vibrational structure, followed by a characterization of the resulting states and a computation of the Franck-Condon profile. This is a formal section that introduces the required protocols for the emulation: mathematical operations; preparation, control, and measurement of a particular experimental setup; and, finally, interpretation of the results. This discussion sets up the requirements that have to be matched by the physical implementation presented in Sec.~\ref{sec:Architecture}.

It is convenient, for practical reasons, to work with Hamiltonians of the form of Eq.~\eqref{eq:Hamiltonian} with full-rank coupling matrices $\hat{A}^{\left(\Theta\right)}$. In order to reduce to zero the dimensionality of the kernel of the coupling matrices, which exists because of the implicit translational and rotational invariance of $A^{\left(\Theta\right)}$, we only need to make the substitution
\begin{equation} \label{eq:nullspace-removal}
\hat{A}^{\left(\Theta\right)} \, \rightarrow \,
\hat{A}^{\left(\Theta\right)} + \sum\limits_j \lambda^2_j \, \vc{\nu}^{}_j \, \vc{\nu}^T_j \, \comma
\end{equation}
where the quantities $\lambda^2_j$ are parameters whose associated frequencies are distinguishable from the physical frequencies of the vibrational model, and the vectors $\left\lbrace\vc{\nu}_j\right\rbrace$ form a basis of its null space. Since the Duschinsky rotations that map different coupling matrices to one another do not affect their kernels, these contributions to the Hamiltonian do not affect the outcome of the experiment, but they effectively remove all zero-mode frequencies from the model and stabilize the following protocols and experiments.

Our protocol is defined as a series of prerequisites, some preprocessing phases, some experimental phases, and, finally, a data-gathering phase.
\\\\
\label{protocol-1}
\emph{Protocol 1 (force-field approach)}.---Let us assume a molecular transition \emph{problem} defined in terms of the following steps:

\begin{itemize}[leftmargin=4.5em]
\item[Step 1:] A set of oscillator masses that do not change throughout the experiment and which form the matrix $M_{jk} = m_j \, \delta_{jk}$,

\item[Step 2:] The initial and final configurations of the vibrational modes defined in terms of the coupling matrices and displacements, $\left\lbrace \hat{A}^{\left(0\right)}, \, \vc{v}^{\left(0\right)} \right\rbrace$ and $\left\lbrace \hat{A}^{\left(f\right)}, \, \vc{v}^{\left(f\right)} \right\rbrace$,

\item[Step 3:] The eigenfrequencies associated with these models, $\left\lbrace \omega_n^{\left(0\right)} \right\rbrace$ and $\left\lbrace \omega_n^{\left(f\right)} \right\rbrace$, upper bounded by $\omega_{\max}$, and an initial state of the molecule, which may be thermal or a ground state, $\rho\left(0\right)$.
\end{itemize}

Let us assume that we have a quantum device, the \emph{emulator}, described by a set of coupled harmonic oscillators,
\begin{equation} \label{eq:emulator-H}
H_E = \frac{1}{2} \, \vc{q}^T \hat{C}^{-1} \vc{q} + \frac{1}{2} \, \vc{\phi}^T \hat{B} \, \vc{\phi}
    - \vc{\phi}^T \vc{V} \comma
\end{equation}
with canonical variables $\left[\phi_j,q_k\right]=i\hbar\,\delta_{jk}$, and with fully adjustable drivings, $V_j$, frequencies, and couplings, $\left\vert B_{ij} \right\vert \in \left[ \, 0, \, B_{max} \, \right]$. 
By comparing Eqs.~\ref{eq:Hamiltonian} and~\ref{eq:emulator-H}, we can map the quantum simulation parameters as in the following steps. 
\begin{itemize}[leftmargin=4.5em]
\item[Step 1:] \label{step-scaling} Compute the following auxiliary rescaled matrices and vectors:
\begin{align}
\label{eq:scaling}
\hat{B}^{\left(0,f\right)} & = \kappa^2 \, \hat{C}^{1/2} \hat{M}^{-\left(1/2\right)}
                                        \, \hat{A}^{\left(0,f\right)}
                                        \, \hat{M}^{-\left(1/2\right)} \, \hat{C}^{1/2} \, \comma
\\ \label{eq:scaling-b}
\vc{V}^{\left(0,f\right)} & = \kappa^{3/2} \, \hat{C}^{1/2}\hat{M}^{-\left(1/2\right)}
                                           \, \hat{A}^{\left(0,f\right)} \, \vc{v} \, \comma
\end{align}
with a possible choice of $\kappa = B_\text{max} / \omega_\text{max} \,$.

\item[Step 2:] \label{step-prepare} Prepare the emulator with the couplings $\hat{B}$ and drivings $\vc{V}$ given by the previous calculation,
\begin{equation}
\hat{B}_\text{start} = \hat{B}^{\left(0\right)} \comma \;
\vc{V}_\text{start} = \vc{V}^{\left(0\right)} - \vc{V}^{\left(f\right)} \, \point
\end{equation}

\item[Step 3:] \label{step-initial} Prepare the initial state in this emulator, which may be either a ground state ($\ket{0,\ldots,0}$) or a (Gaussian) thermal state.

\item[Step 4:] \label{step-final} Abruptly switch, during an appropriate time $T_\text{sw}$, to the final configuration
  \begin{equation}
    \hat{B}_\text{end} = \hat{B}^{\left(f\right)} \comma \; \vc{V}_\text{end} = 0 \, \point
  \end{equation}
  
\item[Step 5:] \label{step-measurement} Measure the \emph{total energy} stored in the entire resonator array, $E$.

\item[Step 6:] \label{step-reconstruct} Based on the previous measurement, gather statistics and reconstruct $P\left(E\right)$, including the uncertainties in the estimation of the probability.
\end{itemize}

As mentioned before, we will discuss the practical aspects of this protocol in a later section, where we explain how to implement the different steps---for instance, model~\eqref{eq:emulator-H} or step 5---using a particular architecture. Before that, however, we need to explain the theoretical considerations behind the protocol, its steps, and the reasoning behind its design.

\subsection{Model scaling}

Our first remark is that a molecule and a superconducting emulator do not share the same energy scale: In general, the energies of a superconducting implementation (roughly in the megahertz-to-gigahertz range) are much smaller than the vibrational energies of a molecule (approximately in the terahertz or mid-IR range). This difference 	implies that, for a proper emulation, all of the parameters in the Hamiltonian have to be rescaled accordingly in a way that takes into account the allowed frequency ranges of the experiment, both from the point of view of the measurement bandwidth of the emulator and from the constraints on the temperatures that can be reached in the emulator.

Step 1 of the protocol accomplishes the task of properly designing the emulator parameters in whatever chosen platform. As discussed in Appendix~\ref{app:rescaling}, the transformation in Eq.~\ref{eq:scaling} is an identity that simply changes the length of the canonical variables. This transformation ººproduces a Hamiltonian whose physics is similar, but in which the frequencies have been rescaled from $\omega_n$ to $\Omega_n = \kappa \, \omega_n$. All of the other observables may be similarly reconstructed:
\begin{align}
\label{eq:mapping}
\mathrm{molecule} & \leftrightarrow \mathrm{emulator} \nonumber \\
    \vc{p}        & = \frac{1}{\sqrt{\kappa}} \, \hat{M}^{1/2} \, \hat{C}^{-\left(1/2\right)} \, \vc{q}
                      \, \comma \\
    \vc{x}        & = \sqrt{\kappa} \, \hat{M}^{-\left(1/2\right)} \, \hat{C}^{1/2} \, \vc{\phi}
                      \, \point \nonumber
\end{align}

Finally, it is important to clarify that the choice of scaling in step 1 is not unique. There may be others that are motivated not by the size of couplings but by the maximum achievable frequencies, the setup constraints, etc.

\subsection{Independent resonators and measurement}

Our second remark is that we have engineered steps 3 and 4 in Protocol 1 to move from generic initial ($\hat{B}_\text{start}$) to final ($\hat{B}_\text{end}$) coupling matrices and, finally, to measure the total energy contained in the resonator array. This is indeed possible, as we argue in Sec.~\ref{sec:Architecture}, yet, in many circumstances, it would be more advantageous to end up at a configuration such that every normal mode is associated with an \emph{independent resonator} through its variables $\left\lbrace q_j \comma \, \phi_j \right\rbrace$. This is a particularly easy thing to accomplish when the ``mass'' matrix $\hat{C}$ in the emulator~\eqref{eq:emulator-H} is diagonal. If this independence between resonators is achieved, we can provide means to inquire the populations of these independent resonators individually---for instance, by coupling qubits or other detectors to each resonator, as illustrated in Fig.~\ref{fig:setup}---thereby obtaining information not just on spectroscopic properties but also on entanglement properties and correlations between modes, or full Wigner function representations.

These techniques are discussed in Sec.~\ref{subsec:Measurement}, requiring only a slight modification of the protocol.
\\\\
\label{proto:diagonal}
Protocol 2 (normal mode approach).---This protocol reproduces the assumptions and steps in Protocol 1, but uses the following steps instead.
\begin{itemize}[leftmargin=4.5em]
\item[Step 1$^\prime$] Compute the following auxiliary rescaled matrices and vectors:
\begin{align}
\label{eq:scaling-2}
B^{\left(0,f\right)} & = \kappa^2 \, \hat{C}^{1/2} \hat{M}^{-\left(1/2\right)} \, \hat{A}^{\left(0,f\right)} \,
                                     \hat{M}^{-\left(1/2\right)} \, \hat{C}^{1/2} \, \comma
\\ \label{eq:scaling-2b}
\vc{V}^{\left(0,f\right)} & = \kappa^{3/2} \, \hat{C}^{1/2} \hat{M}^{-\left(1/2\right)} \,
                                              \hat{A}^{\left(0,f\right)} \, \vc{v} \, \comma
\end{align}
with a possible choice of $\kappa = \Omega^{}_\text{max} / \omega^{}_\text{max}$.

Diagonalize the target configuration $\hat{B}^{\left(f\right)} = \hat{O} \> \Omega^{\left(f\right)} \, \hat{O}^T$ in order to find the final rescaled eigenfrequencies, $\Omega_j^{\left(f\right)} = \kappa \, \omega^{\left(f\right)}_j$, and the orthogonal transformation $\hat{O}$.
\item[Step 2$^\prime$] Prepare the emulator with the parameters
\begin{equation}
\hat{B}_\text{start} = \hat{O}^T \hat{B}^{\left(0\right)} \hat{O} \comma \; \vc{V}_{start}
               = \hat{O}^T \left( \vc{V}^{\left(0\right)} - \vc{V}^{\left(f\right)} \right) \point
\end{equation}
\item[Step 4$^\prime$] Abruptly switch, at an appropriate time $T_\text{sw}$, to the final configuration of the uncoupled resonators,
\begin{equation}
\hat{B}_\text{end} = \hat{\Omega}^{\left(f\right)} \comma \; \vc{V}_\text{end} = 0 \, \point
\end{equation}
\item[Step 5$^\prime$] Measure the \emph{number of phonons} in each of the decoupled resonators, $n_j$, and reconstruct the energy,
\begin{equation}
E = \sum\limits_j \hbar\omega^{\left(f\right)}_j \, n_j \, \point
\end{equation}
\end{itemize}

\subsection{Quench times and errors}
\label{subsubsec:quenching-times}

Our final remark is that step 4 need not be instantaneous to succeed---indeed, there exists nothing instantaneous in real experimental setups, and real quenches always involve a finite amount of time $T_\text{sw}$ (see Sec.~\ref{sec:parameters}). Fortunately, we can approximately solve the dynamics of the state of the emulator during the quenching window. Using time-dependent perturbation theory, we prove analytically in Appendix~\ref{app:switch} that there exists an upper bound to $T_\text{sw}$ such that any quench faster than this time will produce a final state that is approximately unperturbed and as close to the ideal case $T_\text{sw}\to 0$ as we wish.

To be more precise, we have proven the following.

\label{prop:gaussian}
\emph{Proposition 1}.---If we adjust the quench time as
\begin{equation} \label{eq:quenching-bound}
T_\text{sw} = \epsilon \times O \left( \min\left\lbrace
\frac{1}{\Omega^{}_\text{max}} \comma \,
\frac{2}{\vert \hat{C}^{-\left(1/2\right)} \, \vc{V}_\text{start} \vert} 
\right \rbrace \right) \comma
\end{equation}
the final state will approximate the ideal quench $\left( T_\text{sw} \to 0 \right)$ up to errors $O \left(\epsilon\right)$, in (i) the first and second momenta, (ii) the total energy and (iii) the fidelity of the state.

This proposition indicates that, by choosing a quenching time $T_\text{sw}$ that is $1/\epsilon$ times faster than the bound given by Eq.~\eqref{eq:quenching-bound}, the relative errors on the fidelity and other observables of the final state of the system after the quench can be bounded to be, at most, on the order of $1/\epsilon$.
Notice that the switching is constrained not only by the fastest time scale in the Hamiltonian, but also by the initial displacement of the vibrational modes. Our intuition is that, when the minima of the initial and final configurations are far apart, the Wigner function of the quenched state will start far from the origin and will change very rapidly in phase space, with a velocity that is approximately the displacement times $\omega$.

In order to avoid relevant changes in the quantum state during the quenching window, we need $T_\text{sw}$ to scale as the inverse of the average number of photons stored in the resonator in the quenched state, which is the physical interpretation of $\left\vert \, \hat{C}^{-\left(1/2\right)} \, \vc{V}_\text{start} \right\vert$. This is the reason why the null space is forcibly removed in Eq.~\eqref{eq:nullspace-removal}: Modes in the null space of the vibrational Hamiltonian, which correspond to the rigid degrees of freedom (translational and rotational) of the molecule, would have a marginally low frequency in a realistic emulation in a superconducting architecture. As a consequence, these modes could host a large population of photons. This is undesirable, as it would decrease the switching times $T_\text{sw}$ necessary for approaching the ideal quench, as these photons could leak into the cavities that represent vibrational modes through imperfections in the switching off of the tunable couplings.

\section{Physical implementation}
\label{sec:Architecture}

We have introduced the emulation protocol in a formal way. We now discuss how every stage of the emulation, from the initial preparation to the measurement, can be implemented using superconducting circuits.

\subsection{Tunable resonator array}

\begin{figure}[t]
\includegraphics[width=0.825\linewidth]{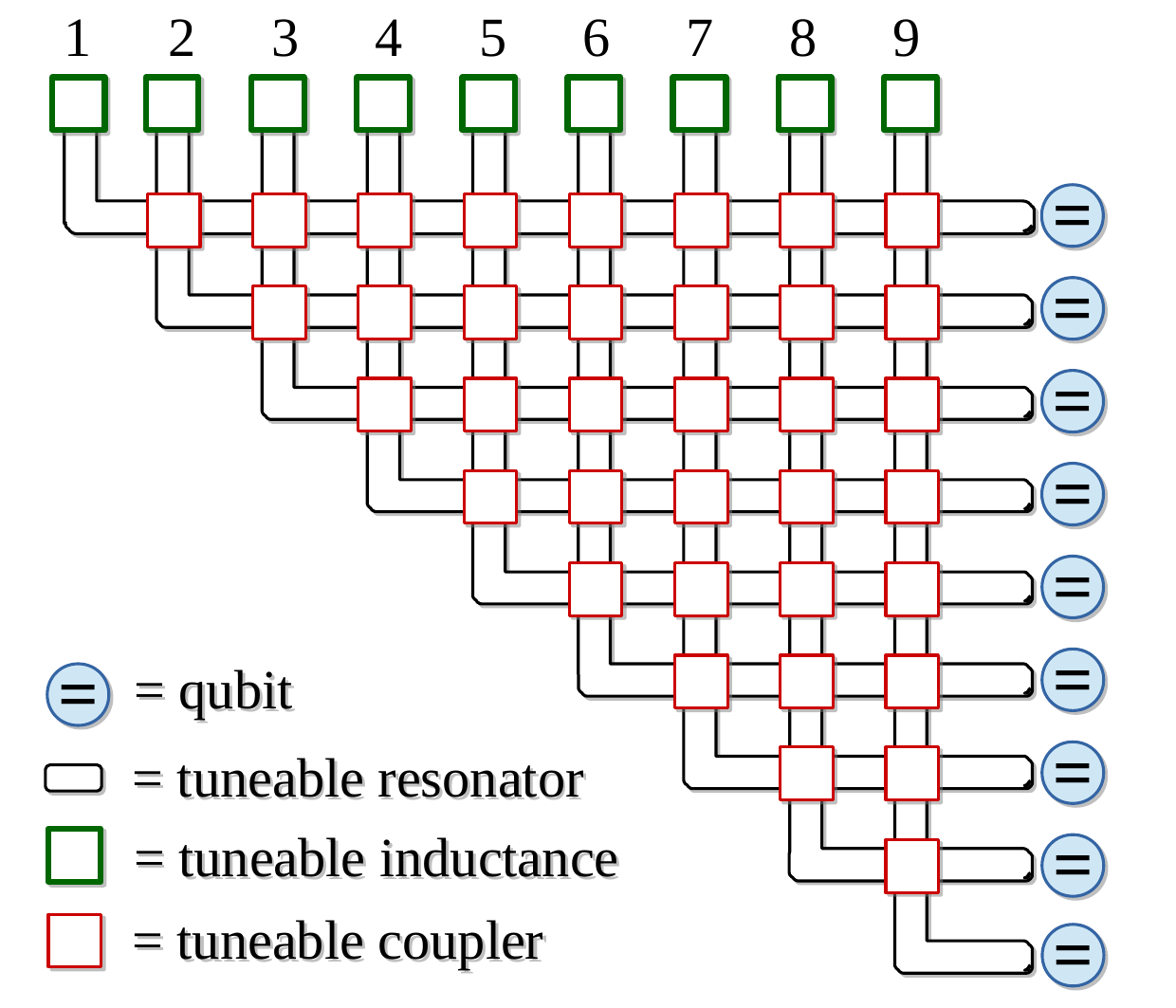}
\caption{Schematic representation of a scalable architecture of superconducting resonators (the black tubes) connected by tunable interactions at their intersections (the red boxes), and with tunable inductors to change their frequencies (the top boxes). Each resonator may or may not be coupled to an additional qubit (the circle) for preparation and measurement purposes.}
\label{fig:setup}
\end{figure}

The basic ingredient in our emulation protocol is the possibility of implementing the model~\eqref{eq:emulator-H} with the tunable parameters $\hat{B}$ and $\vc{V}$. Our suggestion consists in using superconducting microwave resonators for this task. A possible architecture for such an ensemble of resonators is shown in Fig.~\ref{fig:setup}, where we graphically intertwine nine tunable resonators that cross over each other. Note that these resonators interact with each other at their crossing points either directly or through circuits implementing an adjustable mutual inductance.

There exist multiple proposals for implementing both tunable resonators and tunable couplings between them, which rely on different variations of superconducting quantum-interference devices (SQUIDs) for both the tuning~\cite{wilson11,sandberg08} and the coupling~\cite{peropadre13}. We believe, however, that a promising approach is to revisit the D-Wave architecture of flux qubits to implement these types of setups. Specifically, D-Wave qubits, when brought back close to zero flux bias, are nothing but tunable SQUIDs in which the plasma frequency can be adjusted with external magnetic fields. Moreover, there exist robust variations of the SQUID setup that have been proposed and tested with such qubits~\cite{harris09,johnson10}, for the purposes of tuning both frequencies and couplings. In contrast to the qubit regime, our demands for fidelity and dephasing are much more relaxed. The need for less coupled elements, as well as the ongoing progress in the design of controls for superconducting circuits, may significantly improve the switching times for frequencies and couplings, which in the D-Wave architecture were very long.

Irrespective of the architecture that is finally used, the effective lumped element circuit of the resonator array can always be written in the linear form as
\begin{equation}
H = \frac{1}{2} \, \vc{q}^T \hat{C}^{-1} \, \vc{q}
  + \frac{1}{2} \, \vc{\phi}^T \hat{L}^{-1} \, \vc{\phi} + \vc{\phi}^T \hat{L}^{-1}_\text{ext} \, \vc{\phi}^{}_\text{ext}
\, \comma
\end{equation}
where $\hat{C}$ and $\hat{L}$ are, respectively, the capacitance and inductance matrices, and the last term is the inductive energy associated with the coupling with external currents. The terms in the matrix $\hat{C}$ come from the local capacitances of each resonator, $C_{jj}$, as well as the mutual capacitances between neighbors, $C_{j \neq k}\ll C_{jj}$. The inductance matrix contains both the diagonal terms that we use to control the frequency of each resonator, $(\hat{L}^{-1})_{jj}$, and the mutual inductances between different resonators, $(\hat{L}^{-1})_{j \neq k}$.

In this setup, the matrices $\hat{C}$ and $\hat{L}$ and the vector $\vc{I} = \hat{L}^{-1}_\text{ext} \, \vc{\phi}^{}_\text{ext}$ map directly to the equivalent objects $\hat{C}$, $\hat{B}$, and $\vc{V}$ in Eq.~\ref{eq:emulator-H}. All of the inductive elements, $\hat{L}$, and currents, $\vc{I}$, are susceptible to external control when we place tunable elements, such as SQUIDs, either at the ends of the resonators or at the intersections (see. Fig.~\ref{fig:setup}). By using these elements, we can prepare the desired matrices $\hat{B} \sim \hat{L}^{-1}$ and drivings $\vc{V}$.

Note that our formalism for the quantum quenches considers the possibility of having a capacitance matrix with nonlocal interactions, $(\hat{C}^{-1})_{jk} \neq 0$. As discussed before, the only moment at which this matters is at the final stages of Protocol 2, when we attempt to make the matrix $\hat{B}_\text{final}$ almost diagonal (see step 4$^\prime$). If $\hat{C}^{-1}$ is almost diagonal too, it turns out that we have $N$ truly decoupled resonators that can be measured independently using the variant of step 5$^\prime$. However, if there is a residual mutual capacitance $C_{j \neq k}$, the final resonators will be weakly coupled, and there is the possibility of cross-talk when, for instance, we place qubits close to resonance to measure the number of photons (see Fig.~\ref{fig:setup}). Such an effect can be remedied by choosing a different $\hat{B}_\text{final}$ that truly makes the oscillators decouple from each other while the desired final spectrum of the frequencies is preserved.

\subsection{Parameters, frequencies and drivings}
\label{sec:parameters}

As  mentioned in the prerequisites of the protocol, we have to take into account the fact that we may not implement arbitrarily large or small resonator frequencies. In practice, the emulator will work with a spectrum of harmonic eigenfrequencies that are rescaled with respect to the molecular eigenenergies,
\begin{equation}
\Omega^{\left(0\comma f\right)}_n = \kappa \times \omega^{\left(0\comma f\right)}_n \, \point
\end{equation}

We have to choose the emulator scaling $\kappa$ such that the values $\Omega_n$ fit within experimental constraints. For instance, the lower frequency limit is typically imposed by the temperature of the superconducting chip, which will be of the order of tens of mK. This means that, ideally, $\Omega_n \geq \Omega_\text{min} \sim 200$ MHz if we want to start from the ground state of the system, or at least a low-populated state in the least-energetic modes. The upper bound, on the other hand, is given by the Josephson plasma frequencies of the various junctions in the circuitry providing the adjustable frequencies and couplings, reasonably assuming $\Omega_n \leq \Omega_\text{max} \sim$ $20$ GHz for state-of-the-art experiments. In this case, to ensure diabaticity, we have to expect the switching of couplings and frequencies to take place at a rate $1/T_\text{sw} \gg 20$ GHz, which seems to be a feasible figure as shown by circuit QED experiments~\cite{wilson11}.

It is interesting to note that, after the quenching window (bounded by $T_\text{sw}$), the simulation itself is essentially finished and all that remains to be done is to measure the number of photons that the cavities contain. Even short coherence times of tens of nanoseconds, such as those attained in certain quantum annealing architectures comprising a large number of qubits~\cite{bunyk14}, would be sufficient to guarantee the correct reproduction of a diabatic quench, which is the only part of the simulation in which state coherence plays a role. This relaxation on the demands for state coherence may be useful for experimentalists trying to achieve the parameters required to realize a simulation.

Once the range of allowed frequencies is known, we have to adjust the experimental parameters to fit within this region. Considering the specific pair of coupling matrices $\hat{A}^{\left(0\right)}$ and $\hat{A}^{\left(f\right)}$ that is going to be implemented during a simulation, we extract their associated spectra, $\{\omega^{\left(0\comma f\right)}_n\}$. We may now simply take $\omega_\text{max} = \max \, (\lbrace\omega^{\left(0\comma f\right)}_n\rbrace)$ and apply a frequency rescaling,
\begin{equation}
\kappa =  \Omega_\text{max}/\omega_\text{max} \, \point
\end{equation}
Using this scaling, we have a dynamical range
\begin{equation} \label{eq:range}
\frac{\omega_\text{max}}{\omega_\text{min}} < \frac{\Omega_\text{max}}{\Omega_\text{min}}
\sim 100 \, \comma
\end{equation}
which safely lays within the working conditions of most experiments with superconducting circuits. This dynamical range is sufficient for simulating most real molecules, whose typical vibrational bandwidths are typically limited to wave numbers in the range 300--3000 cm$^{-1}$ ($10^{13}$--$10^{14}$ Hz)~\cite{Socrates04,Larkin11,Shimanouchi72,Nakamoto09}. This limitation physically stems from the notion that atoms are coordinated with only so many neighbors and have very weak interactions with distant components of a molecule, and is reflected in the fact that the Duschinsky rotations are almost diagonal for most molecules (see Fig. 2 in~\cite{Dierksen05}).

Finally, the $\lambda_j$ parameters introduced in Eq.~\eqref{eq:nullspace-removal} to prevent an overpopulation of photons in modes with marginally low frequencies (as discussed at the end of Sec.~\ref{subsubsec:quenching-times}) have to be chosen in such a way that, after the rescaling, their frequencies lie in this dynamical range and are distinguishable from the frequencies associated with physical vibrational modes.

\subsection{State preparation}
\label{subsec:StatePreparation}

If the resonator frequencies are engineered according to the requirements laid out in the previous subsection and the dynamical range of the simulated molecule permits it, it is possible to prepare the initial state of the resonators in an almost-zero-temperature state. This can be achieved by simply waiting for a sufficiently long time until the temperature given by the cryostat sinks into the circuit. At temperatures $T_\text{cryo} < \hbar\Omega_\text{min} / K_B$, the number of photonic excitations is negligible. Microwave-induced cooling techniques~\cite{Oliver06,Blais16} may also be considered in flux qubit architectures for zero-temperature simulations requiring the inclusion of modes with frequencies lying below the cryostatic range (below 2 GHz for a 20-mK refigerator) or for faster state preparation~\cite{McClure16}.

However, strictly zero-temperature simulations are not required. Should we wish to prepare a thermal state, it can also be achieved in two possible ways. One is to lower the frequencies of the array so that the effective temperature of the cryostat populates the resonator. In other words, we choose $\kappa$ differently, so that
\begin{equation}
\kappa = \frac{T_\text{molecule}}{T_\text{cryo}}
\; \Leftrightarrow \; \frac{\hbar\Omega_{0}}{K_BT_\text{cryo}} =
                      \frac{\hbar\omega_{0}}{K_BT_\text{molecule}} \, \point
\end{equation}

The range of temperatures available for our simulator are still within a range of computational hardness. Assuming a molecule with a typical bandwidth of $10^{12}$--$10^{14}$ Hz and a circuit operating up to 20 GHz at 20 mK, the circuit would be simulating a molecule with $T_\text{molecule} \sim$ 100 K, which is a moderate temperature with a small occupation per mode. This would still be a classically hard simulation: Low occupation numbers in many---but the least energetic---modes would still require the use of integral-based methods, precluding the use of other approximate techniques~\cite{GarciaPatron17}.

Another possibility would be to couple the resonator array to a source of incoherent microwave radiation with a thermal distribution at the desired effective temperature. Such drivings, which can be obtained from simple resistances, have already been demonstrated in the literature and are routinely used for calibrating tomographic setups for quantum microwaves~\cite{mariantoni10,menzel10}.

\subsection{Measurement}
\label{subsec:Measurement}

The final point of the protocol, after having performed the emulation of the quantum quench, is to extract from the superconducting cavities that emulate the vibrational modes the information necessary to reconstruct the Franck-Condon profile. A variety of strategies may be used for the measurement stages, depending on the physical observables that we intend to characterize. As far as the simulation scheme is concerned, there is not a fundamental reason to select one possibility over the others.

One possible such approach consists of performing a quantum nondemolition (QND) measurement of the number of photons in each of the resonators by using the ancillary qubits depicted in Fig.~\ref{fig:setup}. In the dispersive regime, where qubits are far off-resonant from the resonators, they experience a photon number-dependent energy splitting~\cite{johnson2010quantum,schuster2007}
\begin{equation}
\Delta_{n,j} = \Delta_{0,j} + \left( 2 \, n_j + 1 \right) \frac{g_j^2}{\delta_j} \, \comma
\end{equation}
where $\Delta_{0,j}$ is the bare energy gap of the $j$th qubit, $g_j$ is the coupling to the $j$th resonator and $\delta_j = \Delta_{0,j} - \Omega_j$ is the detuning. The qubits associated with the translational and rotational degrees of freedom may be discarded by keeping them uncoupled from their respective cavities.

Driving each qubit at a different frequency $\Delta_{n,j}$, we detect a Stark shift $\xi_{0,j} = g_j^2 / \delta_j$ on the frequency $\omega^r$ of a readout cavity~\cite{Blais04,Wallraff05}
\begin{equation}
\tilde{\omega}^r_j = \omega^r + \xi_{0,j} \left\langle \sigma^z_j \right\rangle \comma
\end{equation}
which depends on the qubit response to the driving field. If the driving matches the qubit frequency $\Delta_{n,j}$, then $\left\langle \sigma^z_j \right\rangle = 0$; otherwise, $\left\langle \sigma^z_j \right\rangle = -1$. By monitoring transmission through the readout cavity, it is possible to infer the number of photons $n_j$ in the $j$th tunable resonator. When the qubit is driven at the frequency $\Delta_{n,j}$, we measure a cavity frequency $\tilde{\omega}^r_j = \omega^r$, and $\tilde{\omega}^r_j = \omega^r - \xi_{0,j}$ otherwise.

The maximum number of photons that can be resolved using this technique is given by $2\xi_{0,j}/\gamma_j$, where $\gamma_j$ is the decay rate of the $j$th tunable resonator. Using realistic values of circuit QED experiments, we get that $n_\text{max}\simeq 6$~\cite{schuster2007}. This figure is well above the expected number of phonons that populate the vibrational modes of a molecule in spectroscopic experiments ($n_\text{vib} \simeq 3$)~\cite{huh14}.

\section{Anharmonicity}

\begin{figure}[t]
\centering \includegraphics[width=\linewidth]{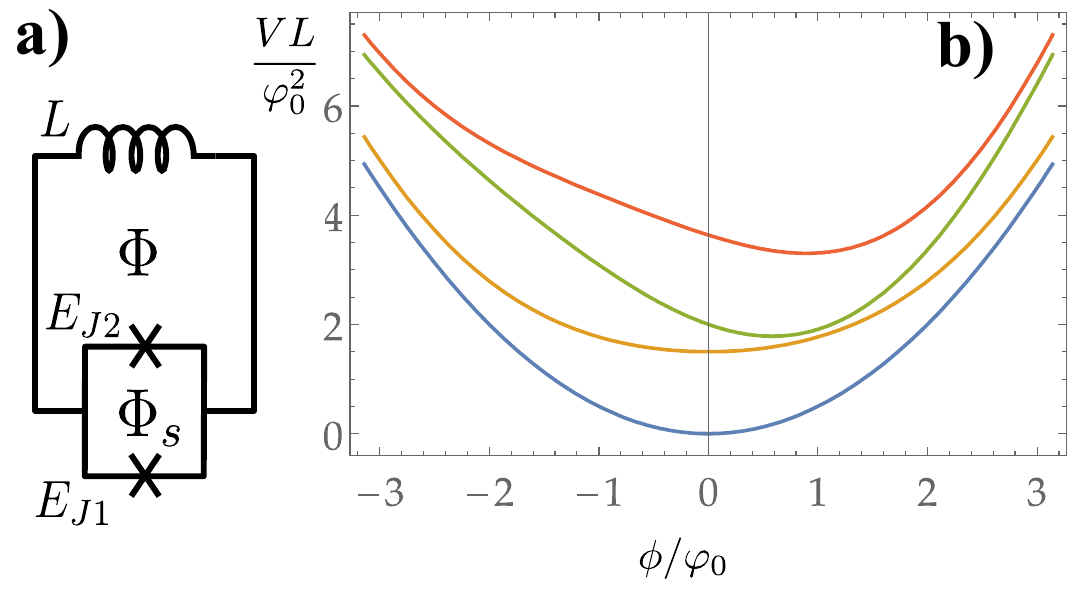}
\caption{(a) Superconducting circuit formed by a couple of Josephson junctions in a SQUID configuration and an associated linear inductance. (b) Energy curves of the circuit potential in Eq.~\ref{eq:squid}, for $\left( L/L_J \comma \, \Phi/\varphi_0 \right) = \left( 0 \comma\, 0 \right) \comma \, \left( 0.5 \comma\, 0 \right) \comma \, \left( 0.7,-\pi/2 \right)$ and $\left( 0.9 \comma\, -\pi/4 \right)$, from bottom to top. Curves have been shifted arbitrarily upwards for better visibility.}
\label{fig:squid}
\end{figure}

One of the most interesting features of the superconducting architecture is the possibility of emulating molecular systems with full control of nonlinear terms. As sketched in Fig.~\ref{fig:the-problem}, general force field potentials are not exactly quadratic near their global minima. A better approximation would be a quartic Taylor expansion around the minimum
\begin{equation} \label{eq:expansion}
V\left( x_0 + \delta{x} \right) \simeq
c_2 \, \delta{x}^2 + c_3 \, \delta{x}^3 + c_4 \, \delta{x}^4
+ O \left( \delta{x}^5 \right) \comma
\end{equation}
where $c_n = \partial^n_x V\left(x_0\right)/n!$. Introducing cubic or quartic terms in classical simulations of molecular quenches is extremely difficult, with molecules having a small number of components already exhausting computational capabilities. This happens because the states involved are no longer Gaussian and cannot be efficiently approximated by a first- and second-order moments (see e.g. Ref.~\cite{huh:2010anharm}). However, adding such nonlinearities to the superconducting setup from Fig.~\ref{fig:setup} is rather straightforward and should be the subject of future work.

As an example, in this section we discuss how replacing a simple $LC$ resonator with the SQUID setup in Fig.~\ref{fig:squid} allows us to achieve any effective nonlinearity with reasonable parameters. We shall work with the inductive energy around the minimum of
\begin{align}
\label{eq:squid}
V\left(\phi\right) = E_J\left(\Phi_s\right) \cos\left[\left(\phi-\Phi\right)/\varphi_0\right]
                   + \frac{1}{2L} \, \phi^2 \, \comma
\end{align}
which contains the static contribution of the linear inductor, $L$, the effective Josephson energy of the SQUID, $E_J$, and the external fluxes trapped in the loops, $\Phi_s$ and $\Phi$. In order to prove \emph{universality} up to fourth order, we only have to find the energy minima of $V\left(\phi\right)$ and verify that it can be expanded like Eq.~\ref{eq:expansion} with any possible ratio of $c_3/c_2$ and $c_4/c_2$.

To achieve these ratios, let us focus on the limit in which the parabolic term $\phi^2/2L$ becomes the dominant contribution, as this can always be achieved by replacing the inductor with a larger junction. The minimum energy configuration may be obtained by expanding around $\phi=0$, which yields the solution $\phi_\text{min} \simeq \Phi/\left(1+L_J/L\right)$, where $L_J\left(\Phi_s\right) = \varphi_0^2/E_J\left(\Phi_s\right)$ is the effective Josephson inductance. Notice that the Taylor expansion around the energy minimum starts at the second order, with $\partial_\phi^2 V \sim 1/L$. The cubic and quartic corrections are found to be
\begin{align}
\frac{c_3}{c_2} & \simeq  \frac{1}{1+L_J/L} \, \frac{\Phi}{\varphi_0^2} \, \comma \\
\frac{c_4}{c_2} & \simeq -\frac{L}{L_J}     \, \frac{1}{\varphi_0^2}    \, \point
\end{align}

Since $L_J \propto \cos\left(\Phi_s\right)$ can be changed in sign and magnitude independently from $\Phi$, it follows that $c_3/c_2$ and $c_4/c_2$ may be tuned separately in order to approximate anharmonic potentials up to order $O\left[\left( \phi - \phi_\text{min} \right)^5\right]$.

The flexibility of this circuit is exemplified in Fig.~\ref{fig:squid} for different values of the SQUID inductance and inner flux. For $\Phi=0$ and $L/L_J=0$, the model is essentially quadratic, and the circuit may be used just like a tunable resonator. However, as we increase the strength of the SQUID, the cubic and quartic terms become dominant and the potential becomes asymmetric, closely resembling the usual Morse potentials.

This idea of anharmonic oscillators can be extended to a multimode circuit and also to the coupling terms between different resonators in Eq.~\eqref{eq:emulator-H}. While scaling up this design to many modes becomes complicated and requires a careful crafting of the different fluxes, the fact is that, as mentioned before, a single-purpose emulator capable of reproducing molecules with a few anharmonic modes would already surpass the computational capabilities of existing classical algorithms. 

As for the rest of considerations in this work, the addition of weak anharmonicities does not significantly modify the discussion on the quench times or the measurements. In the first case, we might have to verify that the quench time is shorter than the inverse of the anharmonic terms, $T_\text{sw} < 1/c_n \left\vert \hat{C}^{-\left(1/2\right)} \, \vc{V} \right\vert^{n/2}$, extending the previous bounds. However, as, in general, $c_{3,4} \gg c_2$, we expect that the introduction of nonlinearities will not impose further constraints in the quantum quench's dynamics. In the case of the measurements, we may resort to spectroscopic means to interrogate the resonator energies, as described in Appendix~\ref{app:measurement}.

\section{Discussion}
\label{sec:outlook}

\subsection{Classical complexity and efficiency}

Before discussing the variety of problems that can be embedded in our quantum simulation scheme and the resulting efficiency, it is illustrative to discuss the current state of the art in Franck-Condon profile calculations. The main notion concerning the classical approaches to the problem of simulating vibronic transitions is that resources scaling exponentially with problem size and bandwidth are required. This stems from the fact that, in order to reproduce the Franck-Condon profile, it is necessary to compute as many integrals as there are ways of distributing $M$ excitations over $N$ normal modes~\cite{Dierksen05}:	
\begin{equation}
\text{No. of integrals} = \left(\begin{array}{c} N + M - 1 \\ M \end{array}\right) \point
\end{equation}

The largest problems analyzed in the relevant literature typically deal with extended molecules of simplified geometries with up to 200 atoms~\cite{Cerezo13}. Even in these situations, approximations that drop contributions to the Franck-Condon profile become a necessity. Still, however, the scalings of memory requirements and computation times in these approaches remain unfavorable with problem size and allowed energies per mode~\cite{Borrelli13,jankowiak:2007}. The situation worsens very rapidly when including anharmonic corrections to the force field~\cite{Huh2011a}.

This is a very similar situation to the one that arises in the context of the classical simulation of the boson sampling problem with initial Fock states~\cite{huh14,BosonSampling15}.

\subsection{Quantum simulability: Size}

Two different issues have to be discussed separately when studying the feasibility of reproducing Franck-Condon profiles in our quantum architecture: (i) the variety of realistic problems that can be simulated, and (ii) the resources consumed by the problems that can be implemented in the architecture.

The first point is subtle: Because of the way that resonators are set to interact with each other, the maximum achievable coupling strength of a cavity connected to $Z$ other cavity modes scales, at most, as $Z^{-1}$. This implies that an architecture with a connectivity of order $Z$ will have a corresponding maximum bandwidth $\omega_\text{max}/\omega_\text{min} \sim 1/Z$, which would seem to restrict the size of the molecules that can be studied using our proposal. However, as was already discussed in Sec.~\ref{sec:parameters}, real molecules also exhibit a limited bandwidth, which is due to the fact that atoms in molecules are coordinated with few neighboring atoms and interact very weakly with distant regions of the system, with both factors contributing favourably to our design.

Note also that existing coupling mechanisms, such as Josephson junctions and SQUIDs, add very large prefactors to the coupling terms~\cite{peropadre13,Baust16}, so that significant coordination numbers $Z$ are available before the $1/Z$ scaling kicks in.

Thus, a realistic superconducting architecture for reproducing Franck-Condon profiles of large molecules would make use of a large number of resonators ($N \gg Z$), with a coordination number $Z$ large enough for embedding a large variety of molecules. For example, carbon atoms in organic molecules may have, at most, four nearest neighbors, which limits the number of higher-order neighbors that may be significantly coupled to them. Typically, from a cursory inspection of the available literature, a value of $Z\sim 10$ seems to be more than sufficient~\cite{Dierksen05}.

\subsection{Quantum efficiency: Time}

The efficiency of our quantum simulation approach is limited mainly by three factors that are, in principle, independent of problem size: (i) The time necessary for the preparation of the initial state, (ii) the time necessary for the measurement of the state after the quenching, and (iii) the number of repetitions required to gather the necessary statistics for reconstructing the Franck-Condon profile.

As discussed in Sec.~\ref{subsec:StatePreparation}, there are two main alternatives for preparing the initial state of the superconducting array: spontaneous relaxation of the photons in the cavity states and microwave-induced active cooling techniques. The former are simple to implement physically, while the latter allow the depletion of modes with frequencies below the cryostatic range~\cite{Oliver06,Blais16} and are faster, with recent experiments having achieved preparation times of the order of tens of nanoseconds~\cite{McClure16}.

Since energy decay lifetimes scale inversely with mode frequency, the lowest possible frequency of approximately 200 MHz (see Sec.~\ref{sec:parameters}) has to be considered in order to obtain a conservative estimation of a state preparation time by spontaneous relaxation. For this figure of 200 MHz, and quality factors of $Q\sim 10^4$, the energy decay rate is on the order of 10 $\mu$s. This figure is insufficient for a realistic, conservative estimation of the preparation time, as it refers to a time at which (at zero temperature) the cavity will have lost a fraction ($1-e^{-1} \sim 0.63$) of its initial stored energy. It is prudent to wait a longer time, of about 100 $\mu$s, for the cavities to fully thermalize.

For the measurement strategy presented in Sec.~\ref{subsec:Measurement}, the smallest among the Stark shifts determines the timescale $\tau_m = \left( \min_j \vert\xi_{0,j}\vert \right)^{-1}$ at which this measurement can be performed. As we are in the dispersive regime $g_j = \chi \delta_j$, for $\chi \ll 1$ and with $\delta_j$ being the detuning, then $\tau_m = 2\pi (\chi^2 \min_j \lbrace \delta_j \rbrace )^{-1}$. Choosing qubit frequencies well below the dynamical range of the resonator frequencies $\Omega_j$, the minimum possible detuning is given by the lower limit of approximately 200 MHz to the dynamical range. Picking $\chi \sim 10^{-2}$ as a reasonably small dispersive coupling parameter, we obtain an estimation for the measurement time of $\tau_m \sim 50$ $\mu$s.

Typical quenching times are on a much faster timescale than the state preparation and measurement times, as the maximum working frequency of the superconducting cavities, which ranges between $200$ MHz and $20$ GHz (as discussed in Sec.~\ref{sec:parameters}) gives an upper bound to the quenching times [see Eq.~\eqref{eq:quenching-bound}]. The quenching, then, will take a time of the order of 0.1 ns at most, thus having a negligible bearing on the simulation efficiency as the other two timescales supersede it for reasonable $Q$ factors. This allows the use of cavities with lower quality factors that are leakier and allow for quicker state preparation.

Finally, the number of sampling repetitions $N_R$ required to reconstruct the Franck-Condon profile with a given precision $\eta^{}_{FC}$ can be found to be, in a worst-case scenario, $N_R \simeq 1 / \eta_{FC}^2$~\cite{huh14}.

Taking into account all of the previous considerations, a useful comparison can be drawn between the typical running times of classical algorithms and the time it would take for a superconducting simulation architecture to gather the necessary statistics. Choosing a target precision $\eta^{}_{FC}\sim 10^{-4}$, of the same order as achievable precisions of current classical algorithms, a number of sampling repetitions of $N_R \sim 10^8$ is obtained. Picking a \emph{very} conservative upper bound $\mbox{(preparation + quench + measurement)} \sim 1$ ms for the time it takes to run a single simulation, we get that the total time necessary to reconstruct a worst-case scenario Franck-Condon profile would be $\sim10^4 s$. This estimation for the necessary simulation time is on the same order of magnitude as the typical running times of classical approaches with approximations~\cite{jankowiak:2007,Borrelli13,Dierksen05,Berger97}.

\subsection{An example: formic acid}

As a practical example of the application of this emulation framework, we present the case of the $S_0 \rightarrow S_1$ vibronic transition between two different vibrational configurations the formic acid molecule (HCOOH), and we show how the parameters to emulate a vibronic transition in a superconducting architecture may be picked.

The particular case of this vibronic transition in the formic acid molecule is a paradigmatic example of an extreme case involving different high-energy vibrational modes that have significant overlaps between them during the transition. A consequence of these overlaps is that, due to the resulting excess of energy in the vibrational modes, the molecule can break down into either CO$_2$ + H$_2$ or CO + H$_2$O after the vibronic transition. Having considered these relevant details of this particular transition, the largest normalized value of the ratio $A_{jk}/\omega_j$ is 0.12. There exist proposals to achieve comparable tuneable couplings in transmission line resonators and flux qubits; in more general, less extreme scenarios than that of the formic acid, lower coupling rates should be sufficient.

The data for the force constants $A^{\left(0,f\right)}_{jk}$ and atomic positions $\left\vert\delta\vc{v}_j\right\vert$ in both configurations have been extracted from Ref.~\cite{jankowiak:2007}. Every physical parameter that characterizes the two vibrational models~\eqref{eq:Hamiltonian} for the two different vibrational configurations of the molecule finds its respective counterpart in the superconducting emulator scheme~\eqref{eq:emulator-H}:

\begin{table}[h]
\begin{tabular}{ccccc}
\hline
\multicolumn{2}{c}{Molecular parameters} & \hspace{0.5em} & \multicolumn{2}{c}{Superconducting emulator} \\
\hline
$M_{jj}$ & $1$--$16$ amu &
         & $0.5$--$8$ pF & $C_{jj}$ \\
$A^{\left(0,f\right)}_{jk}$ & $6.1\cdot 10^{-3}$--$101$ eV$/\angstrom^2$ &
                            & $2.37\cdot 10^{-4}$--$3.83$ nH$^{-1}$ & $B^{\left(0,f\right)}_{jk}$ \\
$\hbar\omega^{}_j$ & $62$--$467$ m$eV$ &
                   & $1.33$--$10$ GHz & $\Omega^{}_j/2\pi$ \\
$\left\vert\delta v_j\right\vert$ & $0.32$--$7.8$ pm &
                                      & $1.27$--$116$ nA & $\left\vert\delta V_j\right\vert$ \\
\hline
\end{tabular}\end{table}

These parameters have been obtained by choosing a maximum working frequency for the resonators in the emulator of 10 GHz, which determines the frequency rescaling factor $\kappa=\Omega_\text{max}/\omega_\text{max}$, and a choice of cavity capacitances consistent with the atomic masses in the molecule. The mutual inductances $B_{jk}$ and drivings $\vc{V}_j$ can be obtained from these previous parameters and the formic acid parameters using~\eqref{eq:scaling} and~\eqref{eq:scaling-b}.

\begin{figure}[t!]
\centering
\includegraphics[ width = 0.85\linewidth ]{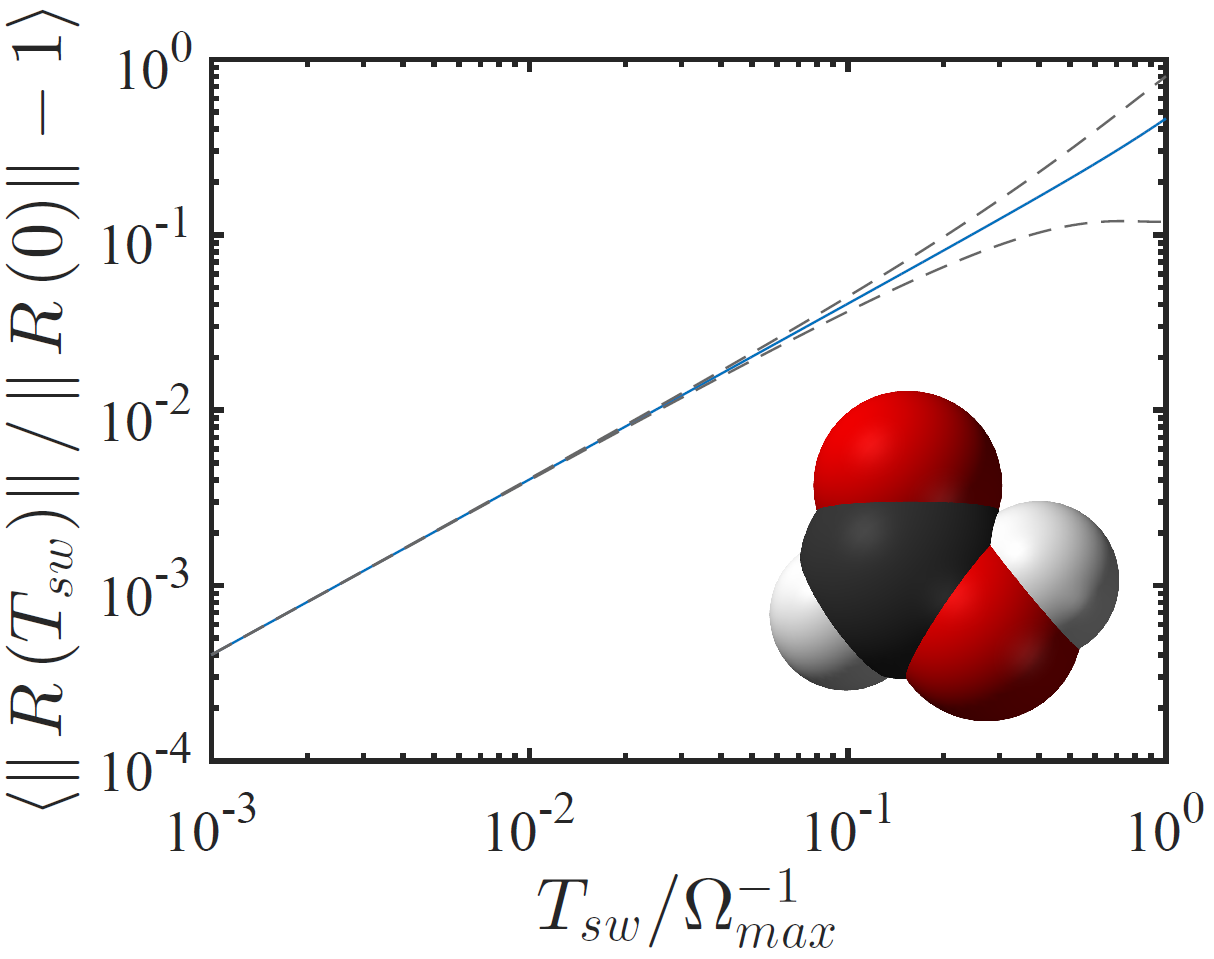}
\caption{Scaling of vector norm differences during a quantum quench of duration $T_\text{sw}$ in the formic acid molecule, estimated from the exact time evolution of the system described by Hamilton's equations from a complete and orthogonal set of initial conditions. It is observed that the mean value of these differences (the solid line) increases linearly with $T_\text{sw}$, which is consistent with the obtained bounds (see Appendix~\ref{app:switch}). The variance from the mean of this set of norm differences (the dashed lines) is shown around the mean value and is found to be small for short times $T_\text{sw} \, \Omega_\text{max} \ll 1$.}
\label{fig:norm-switching}
\end{figure}

In order to reproduce the Franck-Condon profile, the quenching time $T_\text{sw}$ between the two different configurations has to be short enough so that the Franck-Condon approximation (see Sec.~\ref{sec:the-problem}) still applies. As discussed in Sec.~\ref{subsubsec:quenching-times}, the state of the system remains unaffected at the end of quench if $T_\text{sw}$ is shorter than the bound~\eqref{eq:quenching-bound}, which only depends on emulation parameters.

We backup the aforementioned discussion with numerical simulations for different switching times of a realistic quenching process between these two vibrational configurations of the formic acid molecule. In the simulations, we reproduce a quench between the two different quadratic models considered (that we call here $H_1$ and $H_2$), using a linear switching profile $H\left(t\right) = \left(1-t/T_\text{sw}\right) H_1 + \, t/T_\text{sw} \, H_2$. Picking a complete and orthogonal set of initial conditions for the atomic positions and momenta, we ran simulations of the time evolution of the system as governed by Hamilton's equations. The observed changes in the vector norms shown in Fig.~\ref{fig:norm-switching} are small enough and scale linearly with $T_\text{sw}\Omega_\text{max}$ (see Appendix~\ref{app:switch}), so the deviation from the original state of the system may be bounded just as stated in Proposition~\ref{prop:gaussian}.

\subsection{Summary}

In this work, we provide a complete framework for the quantum emulation of a molecular force field using an array of tuneable microwave resonators which leverages the D-Wave design (see Fig.~\ref{fig:setup}). We derive precise protocols for gathering information about molecular transitions, particularly the Franck-Condon spectra, using such a platform---including a detailed discussion of all steps, from the tuning of the emulator to the measurement protocol. Finally, we provide evidence that this architecture may be even more useful when working beyond the quadratic regime.

Our work is an example of another family of useful problems that can be implemented in a setup consisting of superconducting circuits. It would be experimentally relevant and interesting to pursue the design of single-purpose circuits for highly anharmonic molecules with few atoms. Existing blueprints for the D-Wave architecture, as well as ongoing efforts for higher-fidelity quantum annealers with flux qubits, could be leveraged for this task, with the added benefit of offering the possibility of using faster active cooling techniques for state preparation.

\acknowledgements

This work has been supported by MINECO/FEDER Project No. FIS2015-70856-P and CAM PRICYT Research Network QUITEMAD+ S2013/ICE-2801. D. G. O. acknowledges support from MINECO through FPI Grant No. BES-2013-066486. B. P. acknowledges the Air Force of Scientific Research for support under Grant No. FA9550-12-1-0046. J. H. acknowledges support from the Basic Science Research Program through the National Research Foundation of Korea (Grant No. NRF-2015R1A6A3A04059773) and the Mueunjae Institute for Chemistry (MIC) postdoctoral fellowship.

This work has been funded by MINECO Project FIS2015-70856-P and CAM Research Network QUITEMAD+. D.G.O. acknowledges support from MINECO by FPI grant BES-2013-066486. B.P. acknowledges the Air Force of Scientific Research for support under award: FA9550- 12-1-0046. J.H. acknowledges supports by Basic Science Research Program through the National Research Foundation of Korea (NRF-2015R1A6A3A04059773) and Mueunjae Institute for Chemistry (MIC) postdoctoral fellowship.

\appendix

\section{Rescaling of Hamiltonians}
\label{app:rescaling}

Let us assume that we want to simulate a quadratic model
\begin{equation}
H = \frac{1}{2} \, \vc{p}^T \hat{M}^{-1} \vc{p}
  + \frac{1}{2}    \left( \vc{x}-\vc{v} \right)^T \hat{A}_{ij} \, \left( \vc{x}-\vc{v} \right)
\comma
\end{equation}
with canonical variables $\left[x_i \comma\, p_j\right] = i\hbar \, \delta_{ij}$, using a family of tunable Hamiltonians
\begin{equation}
H_E = \frac{1}{2} \, \vc{q}^T \hat{C}^{-1} \vc{q} + \frac{1}{2} \, \vc{\phi}^T \hat{B} \, \vc{\phi}
    - \vc{\phi}^T \, \vc{V} \, \comma
\end{equation}
with alternative canonical variables, $\left[ \phi_i \comma \, q_j \right] = i\hbar \, \delta_{ij}$, and different energy scales. We now show that both models can be mapped to each other through a suitable change of scales. To do so, let us write the transformation
\begin{equation}
\vc{q} = \hat{U}^T \vc{p} \comma \quad \vc{\phi} = \hat{U}^{-1} \vc{x} \, \comma
\end{equation}
which preserves the commutation relations
\begin{align}
\left[ \phi_i \comma\, q_j \right] = \sum\limits_{m,n} \left(\hat{U}^{-1}\right)_{im} U_{nj}
                                     \left[ x_m \comma\, p_n \right] = i\hbar \,\delta_{ij}
\,\point
\end{align}

This leads to the model
\begin{align}
H_E & = \frac{1}{2} \, \vc{p}^T \hat{U} \, \hat{C}^{-1} \hat{U}^T \vc{p}
      + \frac{1}{2} \, \vc{\phi}^T \left(\hat{U}^{-1}\right)^T \hat{B} \, \hat{U}^{-1} \vc{\phi}
      \> + \nonumber \\
      &  - \vc{x}^T \left(\hat{U}^{-1}\right)^T \, \vc{V} \,\point
\end{align}

In order for $H$ and $H_E$ to be equivalent, we simply need $H_E = \kappa H + E_0$, with some constants $\kappa$ and $E_0$. This assumption leads to the condition
\begin{equation}
\kappa \, \hat{M}^{-1} = \hat{U} \, \hat{C}^{-1} \hat{U}^T \comma\; \hat{B} = \kappa \, \hat{U}^T \hat{A} \, \hat{U} \comma\;
\vc{V} = \kappa \, \hat{U}^T \hat{A} \, \vc{v} \, \point
\end{equation}
Using the fact that $\hat{C}$ and $\hat{M}$ are symmetric matrices, we deduce that
\begin{align}
\hat{U}^T & = \sqrt{\kappa} \, \hat{C}^{1/2} \hat{M}^{-1/2} \comma &
\left(\hat{U}^{-1}\right)^T = \frac{ \hat{M}^{1/2} \, \hat{C}^{-1/2} }{\sqrt{\kappa}} \,\comma \\
\hat{U}   &  = \sqrt{\kappa} \, \hat{M}^{-1/2} \hat{C}^{1/2} \comma &
\hat{U}^{-1} = \frac{\hat{C}^{-1/2} \, \hat{M}^{1/2}}{\sqrt{\kappa}} \,\point 
\end{align}
and obtain the suitable oscillator parameters
\begin{align}
\hat{B} & = \kappa^2 \,     \hat{C}^{1/2} \hat{M}^{-1/2} \hat{A} \,
                            \hat{M}^{-1/2} \hat{C}^{1/2} \comma \\
\vc{V}  & = \kappa^{3/2} \, \hat{C}^{1/2} \hat{M}^{-1/2} \hat{A} \, \vc{v}.
\end{align}

At this point we have absolute freedom to choose $\kappa$. We can select
\begin{equation}
\kappa = \frac{B_\text{max}}{\omega_\text{max}} \, \comma
\end{equation}
where $\omega_{\text{max}}$ is the largest eigenfrequency of the normal modes in $H$, and $B_{\text{max}}$ is the largest dynamical range of the eigenfrequencies and couplings in $\hat{B}$;  that is, $B_{\text{max}} = \max_{kl} \left\vert B_{kl} \right\vert$. We assume that there are no restrictions in the strength of $V$.

Finally, notice that the mapping of Hamiltonians is accompanied by a mapping of physical observables, which becomes Eq.~\eqref{eq:mapping} once our choice of $\hat{U}$ is made.

\section{Switching times and diabatic condition}
\label{app:switch}

We can give an upper bound to the time $T_\text{sw}$ required to switch Hamiltonians while preserving the state of the system~\eqref{eq:FC-State} with sufficient fidelity. Without loss of generality, we assume a linear interpolation between the initial and final couplings, which translates into a linear interpolation between Hamiltonians,
\begin{align}
H_E \left(t\right) = & \left( 1 - \frac{t}{T_\text{sw}} \right) H_\text{E,start}
                     + \frac{t}{T_\text{sw}} \, H_\text{E,final} \nonumber \\
= & \frac{1}{2} \, \vc{q}^T C^{-1} \vc{q}
  + \frac{t}{T_\text{sw}} \, \vc{\phi}^T B_\text{final} \, \vc{\phi} \, + \\
  & + \left( 1 - \frac{t}{T_\text{sw}} \right)
      \left[ \, \vc{\phi}^T B_{start,ij} \, \vc{\phi} - \vc{\phi}^T \, \vc{V}_\text{start} \, \right]
\point \nonumber
\end{align}

Our goal is making $T_\text{sw}$ short enough that the state remains almost unperturbed. Since we are interested in the total energy only, it suffices for us to verify that the Heisenberg equations for $\phi_j\left(t\right)$ and $q_j\left(t\right)$ are as close to stationary as possible.

In order to give a proper scale for the proximity of observables and states, we group the canonical operators
\begin{equation}
\vc{R}^T = \left( X_1 \comma \ldots \comma\, X_N \comma \, P_1 \comma \ldots\comma\,P_N \right) \comma
\end{equation}
defined by
\begin{equation}
\vc{\phi} = \hat{C}^{-1/2} \, \vc{X} \, \comma \quad \vc{q} = \hat{C}^{1/2} \, \vc{P} \, \comma
\end{equation}
with the resulting Hamiltonian
\begin{align} \label{eq:linear-switch}
H_E\left(t\right) & = \frac{1}{2} \, \vc{R}^T \hat{D}\left(t\right) \vc{R} - \vc{R}^T \vc{W}\left(t\right) \\
\hat{D}\left(t\right)   & = \left( \begin{matrix} \hat{F}\left(t\right) & 0 \\ 0  & \openone \end{matrix} \right)
\comma \nonumber \\
\hat{F}\left(t\right)   & = \hat{C}^{-1/2} \left[ \left(1-\frac{t}{T_\text{sw}}\right) \hat{B}_\text{start}
                                    + \frac{t}{T_\text{sw}} \, \hat{B}_\text{end} \, \right]\hat{C}^{-1/2}
\comma\nonumber\\
\vc{W}\left(t\right) & = \left( 1 - \frac{t}{T_\text{sw}} \right) \hat{C}^{-1/2} \, \vc{V}_\text{start}
\,\point\nonumber
\end{align}

The evolution equations for the canonical variables become
\begin{equation} \label{eq:HamiltonEquations}
\frac{\diff \vc{R}}{\diff t} = \hat{J} \left[ \hat{D}\left(t\right) \, \vc{R} + \vc{W}\left(t\right)\right] \comma
\end{equation}
where $\hat{J}$ is the matrix of commutators $J_{jk} = \left[ R_j \comma\, R_k \right]$
\begin{equation}
\hat{J} = \left( \begin{matrix} 0 & i \openone \\ -i\openone & 0 \end{matrix} \right) \point
\end{equation}

Our goal is to ensure that $\vc{R}\left(T\right) - \vc{R}\left(0\right)$ is as small as possible. More precisely, we will ensure the following.

\label{prop:switch}
\emph{Proposition 1}.---In our protocol, let us denote by $\Omega_\text{max}$ the largest frequency of the initial or final oscillator configuration. Then, if we switch couplings and frequencies over a time up to an instant
\begin{equation}
T_\text{sw} = \epsilon \times \mathcal{O} \left( \min\left\lbrace
\frac{1}{\Omega_\text{max}} \comma \, \frac{2}{\vert \hat{C}^{-1/2} \, \vc{V}_\text{start} \vert} 
\right \rbrace \right) \comma
\end{equation}
we can ensure that the canonical observables suffer only small corrections,
\begin{equation}
\vc{R} \, \left(T_\text{sw}\right) \sim
\vc{R}_\text{sw}\left(0\right) \times \left[ 1 + \mathcal{O} \left(\epsilon\right) \right]
+ \mathcal{O}(\epsilon).
\end{equation}
\emph{Proof of proposition 1}.---
The formal solution to the Heisenberg dynamics of our observables is given by
\begin{equation}
\label{eq:formal-dynamics}
\vc{R}\left(T\right) = \hat{U}\left(T\comma\,0\right) \vc{R}\left(0\right)
                     + \int_0^T \diff t \, \hat{U}\left(T\comma\,t\right) \vc{W}\left(t\right) \comma
\end{equation}
with the orthogonal operator given by
\begin{equation}
\frac{\diff}{\diff t} \, \hat{U}\left(t\comma\,t_0\right) =
\hat{J} \, \hat{D}\left(t\right) \hat{U}\left(t\comma\,t_0\right)\comma\quad \hat{U}\left(t_0\comma\,t_0\right) =
\openone \, \point
\end{equation}

The second term in Eq.~\ref{eq:formal-dynamics}, which we call $\vc{R}_\text{drive}$, can be easily bounded by 
\begin{equation}
\left\vert\vc{R}_\text{drive} \right\vert \leq
\frac{T_\text{sw}}{2} \left\vert \hat{C}^{-1/2} \, \vc{V}_\text{start} \right\vert \comma
\end{equation}
from which it is obtained that
\begin{equation}
T_\text{sw} =
\mathcal{O} \left(\frac{2\epsilon}{\left\vert \hat{C}^{-1/2} \, \vc{V}_\text{start} \right\vert}\right)
\;\Rightarrow\; \left\vert \vc{R}_\text{drive} \right\vert = \mathcal{O}\left(\epsilon\right) \point
\end{equation}

We focus now on $\hat{U}\left(t\right)$ and on how it deviates from the identity. Our bound for this term relies on the Magnus expansion of the time evolution orthogonal operator
\begin{equation}
\hat{U}\left(t\comma\,0\right) = e^{\hat{\Omega}\left(t\comma\,0\right)} \comma
\end{equation}
The matrix function $\hat{\Omega}\left(t,0\right)$ is constructed as the series expansion $\hat{\Omega}\left(t\comma\,0\right) = \sum_j \hat{\Omega}_j\left(t\comma\,0\right)$, which is called the Magnus expansion. The contributions to this series are obtained recursively from the first term $\hat{\Omega}_1\left(t\comma\,0\right) = \int_0^{t} \diff\tau \, \hat{B} \left(\tau\right)$. The Magnus expansion for $\hat{\Omega}\left(t\comma\,0\right)$ is absolutely convergent~\cite{Blanes09Magnus} if
\begin{equation} \label{eq:MagnusConvergence}
\int_0^{t} \diff\tau \left\Vert \hat{J} \, \hat{D} \left(\tau\right) \right\Vert <
\frac{1}{2} \int_0^{2\pi} \diff x
            \left( 2 + \frac{x}{2} \left( 1 - \cot\frac{x}{2} \right) \right)^{-1}
\comma
\end{equation}
where the rhs integral of Eq.~\eqref{eq:MagnusConvergence} is to be computed numerically. For the specific switching profile~\eqref{eq:linear-switch}, this bound can be approximated as
\begin{equation}
T_\text{sw} \, \frac{\left\Vert \hat{J} \, \hat{D}\left(0\right) \right\Vert +
                     \left\Vert \hat{J} \, \hat{D}\left(T_\text{sw}\right) \right\Vert}{2} < 1 \, \point
\end{equation}
We can now use the fact that $\left\Vert \hat{J}\,\hat{D}\right\Vert=\left\Vert \hat{D}\right\Vert$ and that the spectrum of $\hat{D}\left(t\right)$ gives us the instantaneous eigenfrequencies of the resonator array, $\Omega_n \left(t\right)$. We may thus write
\begin{equation}
T_\text{sw} \max\left\lbrace\Omega_n^{\left(0\right)}\comma\,\Omega_n^{\left(f\right)}\right\rbrace =:
T_\text{sw} \, \Omega_\text{max} < 1 \,\comma
\end{equation}
using the eigenfrequencies of the initial and final problem. If this bound is satisfied, the Magnus expansion may be truncated at first order to estimate that the time evolution during the quenching window differs from the identity as
\begin{equation}
\label{eq:MagnusTruncation}
\left\Vert \hat{U}\left(T_\text{sw}\right) - \openone \right\Vert
\leq \mathcal{O} \left(T_\text{sw} \Omega_\text{max}\right) \point
\end{equation}

The condition for a good fidelity follows from Eq.~\eqref{eq:MagnusTruncation}:
\begin{equation} \label{eq:AbruptCondition}
T_\text{sw} = \epsilon \times \mathcal{O} \left( \frac{1}{\Omega_\text{max}} \right) \point
\end{equation}

This criterion is sufficient to guarantee that the evolution during the switching window does not significantly alter the final energy of the quenched state. If the initial state is Gaussian, which is the case for the ground state of the oscillator or any thermal state, its evolution under the quenching Hamiltonian will preserve this property. We also know that Gaussian states are fully characterized by the first and second momenta of the canonical operators,
\begin{equation}
\vc{r}\left(t\right) := \braket{\vc{R}\left(t\right)} \comma\quad \Gamma_{jk}\left(t\right) :=
\braket{\left\lbrace R_j\left(t\right) \comma\, R_k\left(t\right)\right\rbrace} \point
\end{equation}
Any other expectation value, including the energy and the fidelity of the state relative to the ideal reference, can be computed using these quantities. The bounds from Proposition 1 imply that all moments are well approximated by our choice of quench times
\begin{align}
\vc{r}\left(T_\text{sw}\right) & = \vc{r}_\text{ideal}
                                 + \mathcal{O} \left(\epsilon\right) \comma \\
\Gamma\left(T_\text{sw}\right) & = \Gamma_\text{ideal}
                                 + \mathcal{O} \left(\epsilon\right) \point
\end{align}

We therefore conclude that the final energy distribution and other properties will be well approximated after the quench, as stated in Proposition 1.

\section{Alternative measurement scheme}
\label{app:measurement}

An alternative QND measurement scheme using a setup similar to the one discussed in the main text would consist of the measurement qubits being prepared in the Greenberger-Horne-Zeilinger (GHZ) state $\frac{1}{\sqrt{2}} \left( \ket{00 \ldots 0} + \ket{11 \ldots 1} \right)$. For $\Delta_j \ll \omega_j$, switching on the off-resonant coupling between the qubits and the cavities during a window of time $\tau$ introduces a photon-number-dependent phase factor
\begin{equation}
\phi = \sum\limits_{j=1}^N \frac{g^2_j}{\Omega_j} \, n_j
\end{equation}
into the state of the qubits, which now becomes $\frac{1}{\sqrt{2}} \left( e^{- i \phi \tau} \ket{00 \ldots 0} + e^{ i \phi \tau} \ket{11 \ldots 1} \right)$. If the couplings are tuned such that they correspond to a small (but known) fraction $\chi\ll 1$ of the frequency of the mode to which each qubit is coupled $g_j = \chi\,\Omega_j$ (in order to remain in the dissipative limit), then this phase becomes the total energy of the system up to the multiplicative factor $\chi$:
\begin{equation}
\phi = \chi \sum\limits_{j=1}^N {\Omega_j} \, n_j = \chi E /\hbar \, \point
\end{equation}

Applying the inverse GHZ gate leads to the state
\begin{equation}
\ket{\psi} = \cos\chi E\tau /\hbar \ket{00\ldots 0} - i \sin\chi E\tau /\hbar \ket{11\ldots 1}
\point
\end{equation}

Making use of a qubit readout scheme such as the one discussed in Sec.~\ref{subsec:Measurement}, the qubit (we only need to measure one of them as they are in an entangled state) is found at the excited state with probability $P_1\left(\tau\right)$ as a function of the interaction time $\tau$. This probability corresponds to the probability of the simulation finishing at a certain energy $P\left(E\right)$ through the formula
\begin{equation}
P_1 \left(\tau\right) =
\int \diff E \, P\left(E\right) \left\vert \, \sin\chi E\tau /\hbar\, \right\vert^2 \, \comma
\end{equation}
from which we obtain, by applying the Fourier cosine transformation, that
\begin{equation}
P\left(E\right) = -\frac{4\chi}{\pi \hbar}
\int_0^{\tau_c} \diff \tau \cos 2\chi E\tau /\hbar \, P_1\left(\tau\right) \point
\end{equation}

By repeating many instances of the same simulation for different interaction times $\tau$, the excitation probability $P_1\left(\tau\right)$ is obtained, from which it is possible to obtain the probability $P\left(E\right)$, which is directly connected to the Franck-Condon profile. The Fourier integral is truncated by a cutoff $\tau_c = 2\pi / \Delta\omega$ that is given by the accuracy $\Delta\omega$ with which we seek to reproduce the spectra.

This measurement approach requires a different number of repetitions from the previous one. For a given target precision for the probability of excitation $\Delta P_1$, the number of repetitions necessary to characterize $P_1\left(t\right)$ at a given instant $t$ will be $1/\left(\Delta P_1\right)^2$. Choosing a time resolution $\Delta t$, the total time necessary to gather sufficient statistics to characterize $P_1\left(t\right)$ in the interval $\left[0,\tau_m\right]$ will be
\begin{equation}
 \tau_m \simeq \frac{1}{\left(\Delta P_1\right)^2} \, \frac{\tau_c^2}{2\Delta t} \> \point
\end{equation}

\end{document}